# PAPER CARD-BASED VS APPLICATION-BASED VACCINE CREDENTIALS: A COMPARISON

**Aryan Mahindra**[1], **Anshuman Sharma**[1], **Priyanshi Katiyar**[1], **Rohan Sukumaran**[1], **Ishaan Singh**[1], **Albert Johnson**[1], **Kasia Jakimowicz**[1,2,3], **Akarsh Venkatasubramanian**[1], **Chandan CV**[1], **Shailesh Advani**[1,4], **Rohan Iyer**[1], **Sheshank Shankar**[1], **Saurish Srivastava**[1], **Sethuraman TV**[1], **Abhishek Singh**[2], **Ramesh Raskar**[1,2]

[1]PathCheck Foundation, 02139 Cambridge, USA.
[2]MIT Media Lab, 02139 Cambridge, USA.
[3]Harvard Kennedy School Ash Centre for Democratic Governance and Innovation, 02138 Cambridge, USA.
[4]Terasaki institute for biomedical innovation, University of California, Los Angeles, USA.

raskar@media.mit.edu

## ABSTRACT

In this early draft, we provide an overview on similarities and differences in the implementation of a card-based vaccine credential system and an app-based vaccine credential system. A vaccine credential's primary goal is to regulate entry and ensure safety of individuals within densely packed public locations and workspaces. This is critical for containing the rapid spread of COVID-19 in densely packed public locations since a single individual can infect a large majority of people in a crowd. A vaccine credential can also provide information such as an individual's COVID-19 vaccination history and adverse symptom reaction history to judge their potential impact on the overall health of individuals within densely packed public locations and workspaces. After completing the comparisons, we believe a card-based implementation will benefit regions with less socioeconomic mobility, limited resources, and stagnant administrations. An app-based implementation on the other hand will benefit regions with equitable internet access and lower technological divide. We also believe an interoperable system of both credential systems will work best for regions with enormous working-class populations and dense housing clusters.

## 1 INTRODUCTION

Despite the preventive measures taken to tackle the coronavirus disease 2019 (COVID-19), it has caused unprecedented human deaths (534,099 as of March 17, 2021 CDC (2020)). Over the past year there have been multiple studies looking at the pandemic through digital and clinical lenses(Gandhi et al. (2020b), Gandhi et al. (2020a)). Particularly, the process of testing individuals for COVID-19 has also become an increasingly privacy-invasive process(Morales et al. (2021)).

The introduction of vaccines for the general population creates opportunities to slow down the rapid spread of the virus (News (2021)). However, this opportunity poses new challenges including the strategic, equitable, and privacy-preserving approach of distribution as the world extends mass vaccination drives (EliseoLucero-PrisnoIII (2021)). The current vaccination roll-out has favored highly centralized systems for symptom reporting and vaccine administration. However, a recent paper (Bae et al. (2021b)) proposed a novel solution for decentralized vaccine coordination through the augmentation of CDC's vaccine card with encrypted QR stickers.

In the proposed MIT SafePaths Paper Card (MiSaCa), there are four digitally signed QR code stickers that function as components of a card-based vaccine credential. The distribution mechanism for each, as described in the paper (Bae et al. (2021b)) is as follows - the first QR code is called "coupon" and contains a digital certificate consisting of alpha-numeric codes that are ideally issued by central authorities and get distributed by local governments/employers. The second QR code is termed as "badge" and contains encrypted information such as time, date, and vaccine manufacturer information issued by the vaccination clinic. The third and fourth QR codes are termed as "status" and "passkey" that contain un-encrypted vaccination status and encrypted personal information such as age, name, and sex, respectively. These four coupon codes formulate a workflow that encapsulates the entire user journey for an individual to receive vaccines and show proof of vaccination. After the vaccine recipient receives a coupon QR code, they can enter the coupon's alphanumeric identification into a smartphone application to continue their vaccination journey using app-based vaccine credentials.

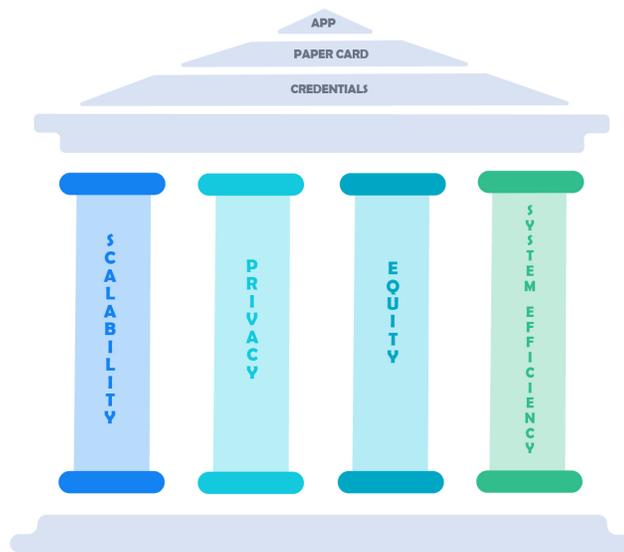

Figure 1: The 4 pillars of vaccine credentials

## 2 METRICS FOR COMPARISONS

This paper compares the two types of vaccine credentials in the domains of fraud and impersonation, adverse event reporting, feasibility, scalability, equity, importability of health data, future data aggregation, and operability to explore each system's features and drawbacks.

### 2.1 FRAUD & IMPERSONATION

Vaccine credentials allow public authorities to verify an individual's vaccination status. This is critical for containing the rapid spread of COVID-19 in densely packed public locations since a single individual can infect a large majority of people in a crowd (Baylor & Marshall (2013)). Jurisdictions can promulgate digital and physical vaccine credentials containing sensitive information such as date of birth, name, and occupation that can be used to extract more sensitive information from an individual. Furthermore, impersonation and fraud of a vaccine credential allow individuals to fake their vaccination status while putting the health of not only themselves, but also that of other individuals at huge risk.

Card-based implementation - A recent statement from the Better Business Bureau exemplifies how scammers in Britain were able to sell fake vaccination cards on digital platforms (like eBay) by extracting personal information from publicly shared vaccination cards (Bureau (2021)). Counterfeit paper cards adhered with QR stickers can be easily created for duplication of vaccine cards and there is a high possibility of identity impersonation when a card is lost or stolen from an individual. Since MiSaCa's QR code stickers are encrypted, an individual's digital certificate and vaccination data is secure from such events.

App-based implementation - Digital credential apps with authentication systems such as one-time passwords (OTPs) and biometric barriers(eg.fingerprints) prevent digitally stored QR codes from duplication. However, there exists a possibility for users to lose their login credentials because of the recent COMB data breach (Bernard Meyer (2021)). This breach has comprised a significant amount of Gmail account's passwords that are often linked to Google's login authentication systems in smartphone applications. Thus, the breach may potentially enable malicious hackers to gain access to the digitally stored vaccine credentials in the app.

## 2.2 ADVERSE EVENT REPORTING

Every publicly available vaccine undergoes excessive testing against a large population of biologically diverse individuals for understanding side-effects, response rates, and complications caused after receiving a vaccination shot in randomized clinical trials (Alaran et al. (2021)). However, the current FDA approved vaccines used to immunize individuals against COVID-19 in the U.S. and globally have been approved at an accelerated pace to curtail the growing mortality and infection rates. This has created an acute need for effective tracking response and adverse reactions to vaccines for the general population. But the methodologies and systems proposed to facilitate such data aggregation using vaccine credentials vary in significant ways.

Card-based implementation - Vaccine recipients are required to report their symptoms by contacting their healthcare provider who uses the Vaccine Adverse Event Reporting System (for Disease Control & Prevention (2020)) to track adverse events from a vaccine recipient in the US. Since this process reveals the person's personal information to the healthcare provider, a card-based implementation provides a new frontier for identifying individuals that prevents exposure of any personally identifiable information. Instead of directly revealing their name, a recipient can provide their vaccination credentials encoded on their coupon sticker on the vaccine paper card as a means of identification for symptom reporting. Because the coupon sticker contains alpha-numeric codes, the vaccine recipient's data is kept secure while they report their symptoms to a healthcare system. However, card-based vaccine credentials require additional interfaces to report adverse events via SMS, web app, or interactive voice response. This creates a need for external infrastructure to be integrated into a card-based vaccine credential system.

App-based implementation - Since apps have the ability to generate noise on servers while registering data, all the information provided can be securely added and transferred across different databases (Bae et al. (2021a)).The ability to anonymize and arrange data without using personally identifiable information is possible with the implementation of differential privacy and secure multiparty computation (Pettai & Laud (2015)). This system, if used by a vaccine recipient makes adverse event reporting through an app extremely privacy preserving. Furthermore, this implementation's approach allows for improved data aggregation and triangulation from various data sources across different geographic locations while functioning as a differential privacy system (Cynthia Dwork (2014)).

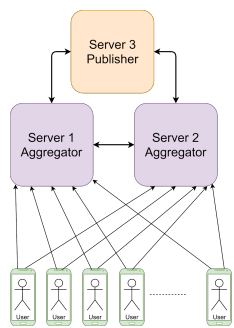

Figure 2: Anonymous adverse event reporting

## 2.3 FEASIBILITY

For the implementation of a vaccine credential system to positively impact the current vaccine rollout, it requires every individual to have a physical or digital vaccine credential. But ensuring that everyone gains access to physical paper cards or software applications on their mobile phone is questionable, given current socio-economic inequalities and digital divides in developed, developing, and underdeveloped countries.

Card-based implementation - In the MiSaCa vaccine credential, the coupon QR code sticker is issued by a jurisdiction while the status, badge, and passkey stickers are issued by vaccination sites. This unidirectional process creates problems for backward compatibility. The lack of multi-channel on-boarding for gaining a card-based vaccine credential complicates the process of distribution in expanding rural and suburban populations. New mechanisms have to be put in place to distribute the coupon, status, badge, and passkey stickers in a more distributed manner. Not only does the new system have to be localized, but also provide more avenues for vaccine recipients to gain a card-based vaccine credential that does not depend on a single jurisdictional authority. Paper-based vaccine credentials also eliminate the need for a smartphone, allowing issuance in multi pen languages and also allow non-personally identifiable information solutions so that private data is not centralized.

App-based implementation - With an app-based vaccine credential system, the vaccine credentials are digitally issued by central authorities and vaccination clinics. This eliminates the friction and chaos of distributing vaccine credentials physically since software applications can digitally transfer and upload data in an efficient and rapid manner. Moreover, the fear of losing QR codes stickers is eliminated as the QR code badges are stored onto the app's permanent storage. However, the challenge concerning adoption of public health-related software applications is extremely difficult to overcome. Without the appropriate allocation of funds to promote and educate communities on using the software correctly, the application will never experience large-scale usage. The lack of technical literacy in communities is another frontier that creates friction in adoption of such public health-related software applications.

## 2.4 SCALABILITY

Challenges for vaccine distribution in the domains of logistics, health outcomes, user-centric impact, and communication are extremely important for performing equitable vaccine distribution(cite). Since vaccine distribution expands its user base by increasing distribution to entire geographic sectors as a whole, the demand for vaccine credentials will accelerate in an exponential manner. This is because entire cohorts of people will be eligible for vaccines at once. The increased demand from different regions will require specific distribution mechanisms to work effectively in a localized manner while targeting different populations in an effective way. Moreover, the state-based vaccine distribution plans vary in significant ways in the US. This causes each state to have entirely separate modes of vaccine credential distribution that can cause scalability issues since the distribution of vaccine, and potentially vaccine credentials, depend solely on the state's limited resources (Freed (2021)). Additionally, the different ways each region prioritize certain segments of their population varies in a significant manner (Chapman et al. (2021)). This compounds the already complex problem of firstly scaling vaccine credential distributions in a region-specific manner, and then delivering vaccine credentials based on a prioritization schema.

Card-based implementation - The MiSaCa paper card containing thestatus, passkey, and badge QR sticker allocated at vaccination sites will be harder to produce with an expanding user base. The ability to increase paper card production with distinct digital certificates will require an airtight system that coordinates sticker productions in new locations effectively. Moreover, the need for all printers to work effectively while producing stickers will be a logistical challenge at all vaccination sites.

App-based implementation - Since the application will always be accessible on the AppStore and Play Store for every user, an expanding user base will not require the system to change in order to meet user demand. However, ensuring that every software update reaches a user is improbable and uncertain, unless the user allows the application to self update within the smartphone. But still, the inertia in making a user sign-up for the application is incredibly high. This is because on-boarding a user through an application would require excessive advertisement and outreach which makes the process slow and expensive.

## 2.5 EQUITY

Vaccine credentials will help communities revitalize their day-to-day operations by allowing businesses and administrations to reopen and function in a scaled manner. But the type of vaccine credential that gets deployed will significantly impact how regular citizens participate in the economy. Without a vaccine credential, individuals will lack a proof of vaccination. And that could result in them being disqualified from joining the workforce. If vaccine credentials are inaccessible, low-income working class individuals in particular will not be able to participate in the reopening workforce. Thus, the choice of a physical or digital vaccine credential impacts a community and its citizens in a significant way.

Card-based implementation - Card-based credentials have been operational for a very long time so vaccinators and the general public are familiar with the concept. . State-issued paper cards are the only way vaccines can be distributed to rural and urban populations while ensuring access to all socio-economic sectors in society. Moreover, card-based vaccine credentials account for the lowest common denominator while ensuring equitable access to vaccines.

App-based implementation - In the US, inaccessibility to stable cellular and broadband internet is excessively large(Linda Poon (2020)). Implementing an app-based vaccine credential on top of this digital divide would further decrease a significant portion of the population from getting vaccines. m, and functional smartphones is an expensive necessity required for app-based vaccine credentials to function properly. However, if app-based vaccine credentials integrate into vaccine distribution software, they can ensure transparent and equitable distribution of vaccines in remote regions (Hasanat et al. (2020)).This is because app-based vaccine credentials create a new frontier for gathering vaccination-related information in a localized manner.

## 2.6 IMPORTABILITY OF HEALTH DATA

The current digital infrastructure within the healthcare industry is highly inflexible and rigid (Elhauge (2010)). This causes information transmission from one system to another to occur slowly. Within the scope of vaccine credentials integrating into this system, the barriers and potential challenges are significantly immense and difficult to overcome. The lack of standardized formatting and generalized entry points makes it harder for a vaccine credential system to fully participate within the current healthcare systems. The main challenge with a vaccine credential is that there is no standard implementation. Because there are no standards, it's hard to imagine vaccine credentials being stored on a centralized healthcare system.

Card-based implementation - Paper cards act as effective end-to-end touch points across the entire vaccination journey. Consequently they can operate with existing legacy systems such as V-safe and VAMS while also entering and transmitting health data using digitally signed QR codes.

App-based implementation - With the implementation of differential privacy in a fully decentralized application, the data stored onto a smartphone will require additional layers of adjustment to integrate with existing legacy systems such as VAMS and V-safe.

## 2.7 FUTURE DATA AGGREGATION

Gathering and assessing health data is of critical importance for deploying vaccines in different geographic sectors. Public health authorities can utilize this aggregated data to target specific geographic regions, deploy vaccines with urgent care to vulnerable populations, and compute the scarcity or abundance of vaccines/vaccination material in different regions. After the deployment and usage of vaccine credentials by the general public, the same credentials can act as access points for gathering data that can be used by public health authorities. This is possible because other than proving an individual's vaccination status, a vaccine credential can record and register an individual's vaccine dose information and symptom history.

Card-based implementation - Since paper cards are issued by central authorities and distributed by employers or regional entities, the collection and triangulation of data will be easier to gather and compute from varying demographic regions with higher specificity. However, the management of garbage data rendered after linking vaccination records to a central authority poses a significant challenge as there is no entity put in place for handling or disposing this type of sensitive data.

App-based implementation - With a direct user-to-issuer pipeline implemented in the app, gathering and computing logistical data would come at the loss of user privacy. However, apps have the ability to generate noise on servers while registering data, all the information provided can be securely added and transferred across different databases (Bae et al. (2021a)).The ability to anonymize and arrange data without using personally identifiable information from a vaccine recipient makes adverse event reporting through an app extremely privacy preserving. Furthermore, this implementation's approach allows for improved data aggregation and triangulation from various data sources across different geographic locations while functioning as a differential privacy system.

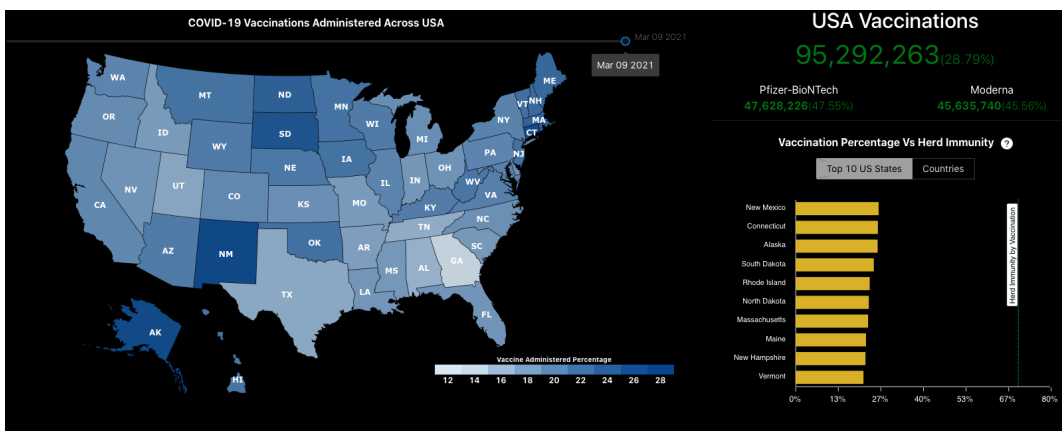

Figure 3: Region-based vaccine data aggregation

## 2.8 OPERABILITY

The vaccine credentials implemented must function properly in all settings to benefit an individual. They must operate with ease while showcasing proof of vaccination without any depreciation of information. The credentials should also showcase data in the easiest manner for a healthcare professional to verify and validate. An app-based vaccine credential and a card-based vaccine credential would achieve this in different ways.

Card-based implementation - Because digitally signed QR codes can function with intermittent connectivity, their implementation does not require additional infrastructure such as cellular data connection and internet access. Moreover, this implementation works for resource constrained environments in an effective and equitable manner.

App-based implementation - With the implementation of a Zero-Knowledge Proof system, parties can validate confidential information without actually actually revealing the information (Bamberger et al. (2021)). The implementation of such cryptographic protocols add to the privacy-preserving operability of an app-based vaccine credential. However, to generate and render additional QR codes into a vaccine recipient's smartphone, intermittent internet and cellular data access would be required. Thus, transferring encrypted QR codes digitally from a vaccine administrator to a vaccine recipient in a cellular dead zone creates barriers for vaccine administration.

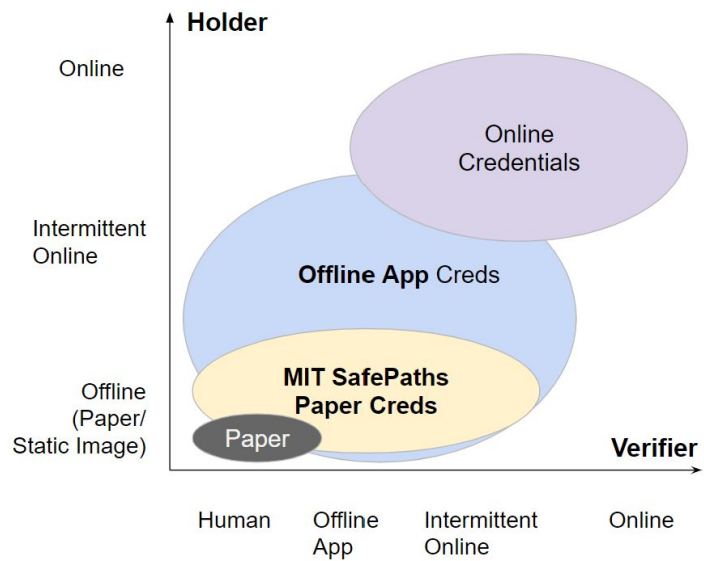

Figure 4: Vaccine credentials working in different environments

## 3 OTHER DIGITAL CREDENTIAL INITIATIVES

### 3.1 VACCINATION CREDENTIAL INITIATIVE (VCI)

VCI is a user-centric initiative program started in collaboration with Microsoft, Oracle, and Mayo Clinic (Initiative (2021)). VCI's goal is to empower individuals with digital access to vaccination records using open and interoperable standards so that an individual can : protect and improve their health, and demonstrate their health status to safely return to travel, work, and school while preserving data privacy.The digital COVID-19 Vaccination Credential Initiative (VCI) makes vaccination records available to patients using an accessible and inter-operable digital format based on globally-accepted standards such as HL7FHIR (HLV.org (2021)). Through VCI, digital vaccination records are provided with a unique QR code that acts as proof for vaccination. These QR codes are stored in health wallets like CommonHealth and AppleHealth and act as a pass for entering public places.

### 3.2 COVID-19 CREDENTIALS INITIATIVE

This decentralized digital credentials initiative is hosted by the Linux Foundation Public Health (Health (2020)). It primarily focuses on gaining the public's trust by implementing an online digital identity called self-sovereign identity. This approach is an alternative to traditional authentication mechanisms through which an individual has unique pivot points for sharing their health records in a digitized manner without the involvement of any 3rd party medium or legacy system integration.

# 4 Summary of comparisons

|   | **Paper card-based implementation** | **Application-based implementation** |
|---|---|---|
| QR stickers | Users have a paper copy of the encrypted QR codes. | Users enter a code in the app to generate encrypted QR codes that get stored into the app digitally. |
| Adverse Event Reporting | Users directly report their symptoms while revealing personally identifiable information to healthcare administrators. | Users report their symptoms anonymously without revealing any personally identifiable information. |
| Fraud | Paper-based credentials will be easier to duplicate because there is a high possibility for impersonation when a paper card is lost or stolen. | Implementation of authentication systems using OTPs and passwords prevent access to digitally stored QR codes. This helps eliminate the possibility of fraud and impersonation. |
| Efficiency | Paper-based cards are easier to carry around because they don't rely on cellular battery and are easy to display. | App generated QR stickers will be hard to display at venues since the uncertainty of cellular batteries will create circumstances where a user cannot access their digitally stored coupon codes. |
| Feasibility | Paper-based cards are difficult to produce and distribute since there is no clear issuance source. A constant and regular supply of materials is required at all vaccination sites to produce paper-based credentials in an efficient manner. | App-generated QR stickers are easy to produce and distribute to every potential vaccine recipient. |
| Scalability | Paper-based card credentials will be harder to safely deliver with an expanding user base. The logistical challenge of distributing paper cards requires additional support from external services for delivery in remote regions. | The process of facilitating user on-boarding with app-based vaccine credentials demands aggressive advertisement and branding. |
| Equity | State-issued paper cards are an efficient way vaccines can be distributed to rural and urban populations while ensuring equitable vaccination access to all socio-economic sectors in society. | The technological divide is quite significant in all countries. Thus, creating an app based solution automatically disqualifies lower socio-economic sectors of society. |

## 5 CONCLUSION

In this early draft, we have evaluated the benefits and potential challenges posed by the implementation of card-based vaccine credential system and an app-based vaccine credential system. The verticals of comparison included domains of symptom reporting, fraud and impersonation, feasibility, scalability, equity, future data aggregation, importability, and operability. Overall, we see the card-based implementation acting as the ideal end point for a user's vaccination journey to function within intermittent offline environments, although the implementation requires multiple systems to be in place for effective QR sticker generation and card distribution. The credentials formulate the lowest common denominator for distributing vaccines in an equitable manner. The app based implementation, on the other hand, allows for quick, securite and efficient transfer of encrypted QR codes, but cannot function in a setting where both the vaccine administrator and user are offline. The implementation of app-based vaccine credentials would require cellular / internet infrastructure to function properly which may create expensive barriers within sectors of the population that have low resources who are trying to re-enter the workforce through proof of vaccination. Some domains favored the implementation of a card-based vaccine credentials system while others favored the implementation of an app-based vaccine credentials system. The impact and choice of solely choosing either approach invites equity and privacy concerns. An implementation of both systems in an interoperable manner will work best for successfully implementing a vaccine credential system.

ACKNOWLEDGMENTS

We are grateful to Riyanka Roy Choudhury, CodeX Fellow, Stanford University, Adam Berrey, CEO of PathCheck Foundation, Dr. Brooke Struck, Research Director at The Decision Lab, Canada, Vinay Gidwaney, Entrepreneur and Advisor, PathCheck Foundation, and Paola Heudebert, co-founder of Blockchain for Human Rights, Alison Tinker, Saswati Soumya, Sunny Manduva, Bhavya Pandey, and Aarathi Prasad for their assistance in discussions, support and guidance in writing of this paper.